\documentstyle[twocolumn,epsf,aps]{revtex}

\draft

\begin{document}

\title{\vspace*{-1.0cm}\hfill {\tt cond-mat/9505107}
       \vspace*{0.5cm}
       \\
	Growth Equation with Conservation Law}

\author{Kent B{\ae}kgaard Lauritsen}

\address{Center for Polymer Studies and Dept.\ of Physics,
	Boston University, Massachusetts 02215~~\cite{email}}

%\date{\today}
\address{\em (\today)}

%\maketitle

%\begin{abstract}
\address{
\centering{
\medskip\em
\begin{minipage}{14cm}
{}~~~A growth equation with a generalized conservation law
characterized by an integral kernel is introduced.
The equation contains the Kardar-Parisi-Zhang,
Sun-Guo-Grant, and Molecular-Beam Epitaxy growth equations
as special cases and allows for a unified investigation of
growth equations. From a dynamic renormalization-group analysis
critical exponents and universality classes are determined for
growth models with a conservation law.
\end{minipage}
}}
%\end{abstract}

\maketitle

%\pacs{PACS numbers: 68.35.Rh, 64.60.Ht, 05.40.+j, 05.70.Ln}
%
% 64.60.Ht = General Studies of phase transitions:
%            Dynamic critical phenomena
% 68.35.Rh = Solid surfaces and solid-solid interfaces:
%            Phase trans. & crit. phenomena
% 05.40.+j = Fluctuation phenomena, random processes, and Brownian motion
% 68.35.-p = Solid surfaces and solid-solid interfaces
%            68.35.Fx  Diffusion; interface formation
%            68.35.Rh  Phase transitions and critical phenomena
% 05.70.Ln = Nonequilibrium thermodynamics

\narrowtext

\section{Introduction}

In order to describe the dynamics and scaling of interface growth,
Kardar, Parisi and Zhang proposed a Langevin equation now
known as the KPZ equation \cite{kpz:1986}.
This equation is constructed to
describe the long-time long-wavelength (hydrodynamic) limit of
the dynamics of nonequilibrium interface growth processes.
The KPZ equation has been studied intensively by analytical and
numerical methods and a number of theoretical results have been obtained
\cite{krug-spohn:1991,family-vicsek:1991,%DO NOT REMOVE THE "%"
hh-zhang:1995,barabasi-stanley:1995}.
Specifically, the scaling of the correlation function
$C(x,t) = x^{2\alpha} \overline{g}(t/x^z)$
characterized by the roughness exponent $\alpha$
and the dynamic exponent $z$ has been established.
However, only poor agreement is obtained between the theoretical results and
experiments \cite{hh-zhang:1995,barabasi-stanley:1995}.
As a result, various modifications of the KPZ equation have been analyzed.

In the present paper we introduce a new growth equation with a generalized
conservation law described by an integral kernel. We will
refer to the equation as the growth kernel equation (GKE).
It contains the KPZ equation as a special case.
In addition, the previously studied Sun-Guo-Grant (SGG) \cite{sgg:1989}
and Molecular-Beam-Epitaxy (MBE) \cite{lai-dassarma:1991}
equations are also contained in our general equation.

The motivation for introducing the GKE equation is to gain information on how
conservation laws change the universality classes
for nonequilibrium growth models, and to allow for a {\em unified\/}
description of growth models studied so far.
Furthermore, one can speculate whether some growth
experiments, which yield exponents that do not agree with the KPZ
exponents, may contain nonlocal growth effects
such as, e.g., the experiments on electrochemical deposition reported in
Refs.\
\cite{kahanda-etal:1992,iwamoto-etal:1994}.

Previous studies of nonlocal terms in interface related topics include
fluctuating lines in quenched random environments.
Domain walls subject to quenched long-range correlated
impurities were studied in \cite{kardar:1987}.
More recently, the dynamic relaxation of drifting polymers
\cite{ertas-kardar:1992-93}
and the critical dynamics of contact line depinning
\cite{ertas-kardar:1994}
were studied, where in both cases the equation of motion includes
nonlocal interaction terms.

\section{Growth Kernel Equation}

The GKE equation for a $d$ dimensional interface $h(x,t)$ reads
\begin{equation}
        \frac{\partial h}{\partial t} =
                \int d^dx' \, K(x-x') \left( \nu \nabla'^2 h +
                        \frac{\lambda}{2} (\nabla' h)^2 \right) + \eta ,
                                                \label{eq:gke}
\end{equation}
with an additive noise $\eta(x,t)$ whose correlations will be specified below.
The kernel $K(r)$ describes nonlocal interactions in the system
\cite{bray:1990}.
We want the total height $H(t)=\int d^dx h(x,t)$ to be
conserved and impose the constraint
$ \int d^dx' ~ K(x-x') = 0 $, which leads to
$\partial H / \partial t  = 0$, provided the noise is chosen to
satisfy $\eta(k=0,t)=0$ (see Eq.~(\ref{eq:gke-noise}) below).
Consequently, the GKE equation conserves the quantity $H(t)$.

The kernel has the behavior
\begin{equation}
        K(r) \sim \frac{1}{r^{d+\sigma}} \mbox{~~~for~~~} r \to \infty,
                                        \label{eq:gke-K}
\end{equation}
where we have introduced an exponent $\sigma$ describing the
long-distance decay. By Fourier transforming, $\sigma=0$ corresponds to
the kernel being a Dirac delta function, and therefore the usual
KPZ equation \cite{kpz:1986}.

The case $K(x-x')=-\nabla_{x}^{2} \, \delta^d(x-x')$
yields the dynamics of the
SGG and MBE equations \cite{sgg:1989,lai-dassarma:1991},
and corresponds to $\sigma=2$.
In order to incorporate these equations,
we introduce a kernel $N(r)$ in the noise correlator
\begin{equation}
        \langle \eta(x,t) \eta(x',t') \rangle = 2D \, N(x-x') \delta(t-t')
                                        \label{eq:gke-noise}
\end{equation}
with the form
\begin{equation}
        N(r) \sim \frac{1}{r^{d+\tau}} \mbox{~~~for~~~} r \to \infty .
                                        \label{eq:gke-N}
\end{equation}
Here, $\tau$ is an exponent independent of $\sigma$.
The case \mbox{$\tau=0$} means no correlations in the noise
(KPZ, MBE) whereas $\tau=2$ corresponds to conserved noise as
it appears in, e.g., the SGG equation.
The correlator (\ref{eq:gke-noise}) implies that $\eta(k=0,t)=0$,
as required above.
We can imagine the noise related to a white noise $\xi(x,t)$ through
$\eta(x,t) =  \int d^dx' ~ R(x-x') \, \xi(x',t)$,
where the kernel $R(r)$ has the large argument behavior
\mbox{$R(r) \sim 1/r^{d+\tau/2}$}.
This form leads to the correlations in
Eqs.~(\ref{eq:gke-noise}) and (\ref{eq:gke-N}).

Under the rescaling $x\to x'=x/b$ the parameters in the
GKE equation change as
\begin{eqnarray}
	\nu     &\to& \nu' = b^{z-2-\sigma} \nu  ,     \\
	\lambda &\to& \lambda' = b^{z+\alpha-2-\sigma} \lambda ,  \\
	D       &\to&  D' = b^{z-2\alpha-d-\tau} D    .
							\label{eq:gke-param}
\end{eqnarray}
For $\lambda=0$, the equation is made scale invariant for
\begin{equation}
	z_{0}^{} = 2+\sigma
		\mbox{~~~~~~and~~~~~~}
	\alpha_{0}^{} = \frac{2+\sigma-d-\tau}{2}  .
					\label{eq:gke-lin-exp}
\end{equation}
If we use these values in the rescaling for $\lambda$ we obtain that
$\lambda'=b^{(2+\sigma-d-\tau)/2} \lambda$,
so naively we expect the critical dimension of the model
to be given by \mbox{$d_c = 2+\sigma - \tau$}.
Therefore, for $d>d_c$ the $\lambda$ term will scale to zero,
whereas for $d<d_c$ the $\lambda$ term will be relevant and
the scaling behavior of the GKE equation will no longer
be described by the naive exponents.
Now we will carry out a dynamic renormalization group (RG) analysis in order
to determine the scaling behavior of the GKE equation.

\section{Renormalization Group Analysis}

We Fourier transform the GKE equation using the rules
for Fourier transformation of convolutions and products
and obtain in the hydrodynamic limit $k \to 0$
\begin{eqnarray}
        h(k,\omega) &=& {G_0(k,\omega)} \eta(k,\omega)
                        -{\lambda \over 2} \, {G_0(k,\omega)}\, k^\sigma\!
			\int^\Lambda \frac{d^dq}{(2\pi)^d}
                                                       \nonumber  \\
                &\times&\!\!
                        \int_{-\infty}^\infty \frac{d\Omega}{2\pi}~
                  q \cdot (k-q)\, h(q,\Omega)\, h(k-q,\omega-\Omega) ,
                                                            \label{eq:gke-h}
\end{eqnarray}
where $G_0(k,\omega)$ is the (bare) propagator defined by the expression
$
	G_0(k,\omega) = { 1 / (\nu k^{2+\sigma} - i \omega) } .
$
$\Lambda$ is the momentum cutoff.
The noise in Fourier space takes the form
\begin{equation}
	\langle\eta(k,\omega)\eta({k}',\omega')\rangle
		= 2D \, k^\tau \, (2\pi)^{d+1}
			\delta^d(k+k') \delta(\omega+\omega').
						      \label{eq:gke-noise-ft2}
\end{equation}

\begin{figure}[htb]
\centerline{
	\epsfxsize=8.0cm
	\epsfbox{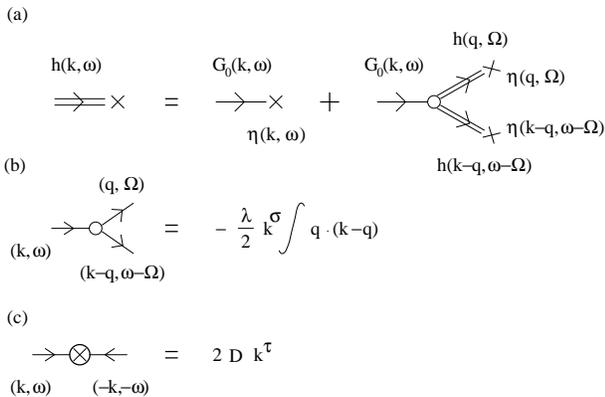}
        \vspace*{0.5cm}
	}
\caption{(a) Diagrammatic representation of the GKE
	     equation~(\protect\ref{eq:gke-h}).
        (b) The vertex $\lambda$ which includes integration over
	    $(q,\Omega)$.
	    The $q \cdot (k-q)$ is associated with the outgoing momenta;
	    $\int \equiv\int^\Lambda\frac{d^dq}{(2\pi)^d}
		\frac{d\Omega}{2\pi}$.
	(c) The contracted noise $2Dk^\tau$, from
            Eq.~(\protect\ref{eq:gke-noise-ft2}).
	}
\label{fig:diagrams}
\end{figure}

The renormalization group consists of coarse-graining
followed by rescaling \cite{ma:1976}.
Coarse-graining: Modes with momenta $e^{-\ell} < k < 1$ ($\Lambda\equiv1$)
are eliminated from the equation of motion.
Rescaling: Wavevectors are rescaled according to
$k \to k'=bk$, with $b=e^\ell$.
The RG procedure is most efficiently carried out by the means of diagrams,
i.e., we represent the GKE equation (\ref{eq:gke-h}) as shown in
Fig.~\ref{fig:diagrams},
cf.\ Refs.\ \cite{kpz:1986,medina-etal:1989,fns:1977}.

After a standard but lengthy calculation one obtains the
one-loop RG flow for the GKE equation \cite{kbl-thesis:1994}
\begin{eqnarray}
        \frac{d\nu}{d\ell} &=& \nu \left( z-2 - \sigma + K_d  \,
            		    {\lambda^2 D\over\nu^3} \frac{2+2\sigma-\tau-d}{4d}
                		\right)  ,
						\label{eq:nu-flow} \\
        \frac{d\lambda}{d\ell} &=& \lambda \left( \alpha + z-2 -\sigma \right),
						\label{eq:la-flow} \\
	\frac{d D}{d\ell} &=& D \left( z-2\alpha-d-\tau + A_{\tau,\sigma} \,
		                       \frac{K_d}{4}\,{\lambda^2 D \over \nu^3}
				 \right) , \label{eq:D-flow}
\end{eqnarray}
where
	$A_{\tau,\sigma} = 1$ for $\tau \ge 2\sigma$,
and
	$A_{\tau,\sigma} = 0$ for $\tau < 2\sigma$.
The geometrical factor \mbox{$K_d = S_d/(2\pi)^d$}, and
\mbox{$S_d = 2\pi^{d/2}/\Gamma(d/2)$}
is the surface area of the $d$ dimensional unit sphere.
Furthermore, the RG flow for the case $\tau \ge 2\sigma$
reduces to the case $\tau=2\sigma$ \cite{kbl-thesis:1994}.
Due to the fact that $\lambda$ does not renormalize to one-loop order,
cf.\ Eq.~(\ref{eq:la-flow}),
one obtains the exponent identity
\begin{equation}
	\alpha + z = 2 + \sigma  .
				\label{eq:exp-id-for-gke}
\end{equation}
The critical exponents are determined from
${d\nu}/{d\ell} =0$ and ${d D}/{d\ell} = 0$, i.e., by fixing $\nu$ and $D$.
It is convenient to introduce the coupling constant
\begin{equation}
	g = g(\ell) = \frac{K_d}{4d} \frac{\lambda^2 D}{\nu^3}
\end{equation}
with the dimension $[L]^{d+\tau-\sigma-2}$.
The RG flow of $g$ for fixed $\nu$ and $D$ becomes
\begin{eqnarray}
	\frac{dg}{d\ell}
		&=& 2g\,\frac{1}{\lambda} \frac{d \lambda}{d\ell} %\nonumber\\
		=
		 2g ( \alpha + z -\sigma -2)
						\label{eq:g-flow}   \\
		&=&
		 (2+\sigma-\tau-d)g-[3(2+2\sigma-\tau-d)-dA_{\tau,\sigma}] g^2.
							\nonumber
\end{eqnarray}
The fixed points (FP) for $g$ are
\begin{equation}
	g_{0}^{*} = 0,~~~~
	g^* = \frac{2+\sigma-\tau-d}{3(2+2\sigma-\tau-d) - dA_{\tau,\sigma}} .
					\label{eq:gke-fp}
\end{equation}
Physically, we have that $\nu, D > 0$. As a result,
FPs where \mbox{$g^* < 0$} are unphysical since they would lead to
an imaginary value for $\lambda$.

\section{Results}

\begin{figure}[htb]
\centerline{
	\epsfxsize=8.0cm
	\epsfbox{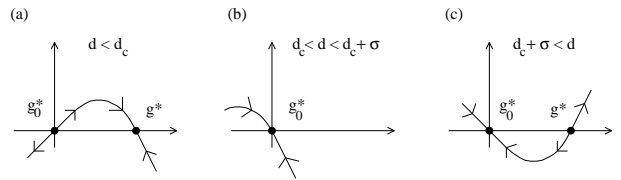}
	%\vspace*{0.5cm}
	}
\caption{Coupling constant flow for $\tau < 2\sigma$.
	}
\label{fig:fp-gke}
\end{figure}

First, we discuss the case $\tau < 2\sigma$.
The critical dimension is $d_c=2+\sigma-\tau$.
In Fig.~\ref{fig:fp-gke} we show the RG flow of the coupling constant $g$,
for various dimensions.
For $d<d_c$ the trivial FP $g_{0}^{*}=0$ is unstable,
whereas $g^*$ is stable.
For $d>d_c$, $g_{0}^{*}$ becomes stable.
For $d>d_c+\sigma$, $g^*$ is again positive but now an unstable FP.
Probably this latter behavior is
an artifact due to the fact that the FP is only known
from the one-loop expansion \cite{kbl-thesis:1994}.

The FPs and exponents can be calculated in an \mbox{$\epsilon = d_c -d$}
expansion.
For the trivial FP $g_{0}^{*}=0$ we recover the
exponents for the linear GKE equation, cf.\ Eq.~(\ref{eq:gke-lin-exp}).
Consequently, they will describe the GKE system for
dimensions $d>d_c$, where the trivial FP is stable,
cf.\ Fig.~\ref{fig:fp-gke}(b).
The nontrivial FP is to first order in $\epsilon$ equal to
\mbox{$
	g^*= \epsilon/3\sigma ,
$}
with the exponents
\begin{equation}
	\alpha = \frac{\epsilon}{3}=\frac{2+\sigma-\tau-d}{3},
					\label{eq:gke-exp-a}
\end{equation}
and
\begin{equation}
	z = 2 + \sigma -\frac{\epsilon}{3} = \frac{d+\tau+2\sigma+4}{3} .
					\label{eq:gke-exp-z}
\end{equation}
These values are the exponents for the GKE equation
in dimensions $d<d_c$,
and are exact to all orders in $\epsilon$ due to the
non-renormalization of $\lambda$ and $D$ in the case $\tau<2\sigma$,
cf.\ Eqs.~(\ref{eq:la-flow}) and (\ref{eq:D-flow}).

Next, we discuss the case $\tau \ge 2\sigma$.
We again remark that for any $\tau > 2\sigma$ we get the
behavior for $\tau = 2\sigma$, and therefore we only have
to discuss the latter case \cite{kbl-thesis:1994}.
The critical dimension is $d_c=2-\sigma$.
With $\epsilon=d_c-d$, the nontrivial FP becomes for $\sigma \neq \frac{1}{2}$
\begin{equation}
	{g^*} = \frac{\epsilon}{2(2\epsilon+2\sigma-1)}
		= \frac{\epsilon}{2(2\sigma-1)} + O(\epsilon^2)   .
						\label{eq:g*-eps}
\end{equation}

\begin{figure}[htb]
\centerline{
	\epsfxsize=8.0cm
	\epsfbox{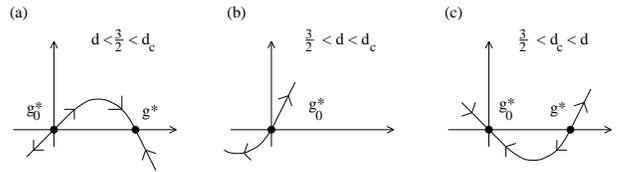}
	%\vspace*{0.5cm}
	}
\caption{Coupling constant flow for $\tau \ge 2\sigma$,
	and $\sigma \le \frac{1}{2}$.
	For dimensions $d>d_c$ (c) shows the possibility of
	a phase transition at $g^*$ (see the text).
	}
\label{fig:fp-kpz}
\end{figure}

The case \mbox{$0 \le \sigma \le \frac{1}{2}$}:
Here, $g^*$ is negative,
and a FP expansion in powers of $\epsilon$ does not exist.
The KPZ equation is a well-known example of this failure of the
$\epsilon$ expansion.
The two-loop results for the KPZ equation also show the failure of the
$\epsilon$ expansion around the critical dimension $d_c=2$
\cite{frey-tauber:1994}.
In Fig.~\ref{fig:fp-kpz} we show the RG flow for $g$ in the
case where $\sigma < \frac{1}{2}$.

In order to obtain the exponents
we can use the one-loop result (\ref{eq:gke-fp})
of the $g^*$ fixed point.
For the KPZ equation this gives the exact exponents in $d=1$
(cf.\ \cite{kpz:1986}),
but despite this ``success'' the method is uncontrolled due to the fact
that $g^*$, or $\lambda^*$, is
not small at the FP, which has been the underlying assumption under the
whole RG calculation and series expansion in $\lambda$.
The direct substitution of (\ref{eq:gke-fp})
into the expressions for the exponents results in the values
(with $\tau=2\sigma$)
\begin{equation}
	\alpha =  \frac{(2-d)(2-\sigma-d)}{2(3-2d)}   ,
					\label{eq:a-direct}
\end{equation}
and
\begin{equation}
	z = 2 + \sigma - \frac{(2-d)(2-\sigma-d)}{2(3-2d)}   ,
					\label{eq:z-direct}
\end{equation}
which in $d=1$ yields $\alpha = (1-\sigma)/2$ and $z=3(1+\sigma)/2$.
For $\sigma=0$ this reduces to the KPZ exponents \cite{kpz:1986}.

\begin{figure}[htb]
\centerline{
	\epsfxsize=8.0cm
	\epsfbox{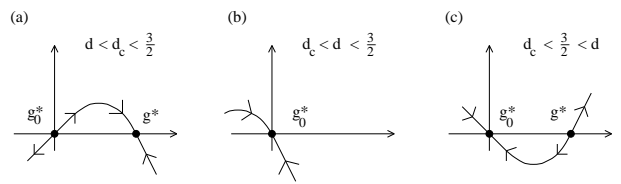}
	%\vspace*{0.5cm}
	}
\caption{Coupling constant flow for $\tau \ge 2\sigma$,
	and $\sigma > \frac{1}{2}$.
	}
\label{fig:fp-new}
\end{figure}

The case \mbox{$\sigma > \frac{1}{2}$}:
In Fig.~\ref{fig:fp-new} we show the RG flow for $g$ in this case.
For dimensions $d>\frac{3}{2}>d_c$ the $g^*$ fixed point is positive;
probably this is an artifact due to the one-loop result.
Now the $\epsilon$ expansion is possible.
We can obtain the exponents to first order in $\epsilon$
at the $O(\epsilon)$ fixed point (\ref{eq:g*-eps}) with the result
\begin{equation}
	\alpha  = \frac{\sigma}{2(2\sigma-1)} \epsilon + O(\epsilon^2)  ,
					\label{eq:a-eps}
\end{equation}
and
\begin{equation}
	z = 2 + \sigma - \frac{\sigma}{2(2\sigma-1)} \epsilon + O(\epsilon^2) .
					\label{eq:z-eps}
\end{equation}

For $\tau=2\sigma$ ($\tau \ge 2\sigma$) there is the possibility of
carrying out an expansion in the quantity $\epsilon'=d-d_c = d+\sigma-2$
for dimensions above $d_c$.
The FP is
\begin{equation}
	g^*(\epsilon') = \frac{1}{2(2\epsilon'+1-2\sigma)} \epsilon'
			= \frac{1}{2(1-2\sigma)} \epsilon' + O(\epsilon'^2)
				\label{eq:rpt-fp}
\end{equation}
and it corresponds to a phase transition
in the model, cf.\ Fig.~\ref{fig:fp-kpz}(c).
The phase transition is between the $g_{0}^{*}$ ``smooth'' phase and
the strong-coupling rough phase.
When $\sigma>\frac{1}{2}$ the $g^*$ fixed point is negative,
i.e., the $\epsilon'$ expansion is only meaningful for $\sigma<\frac{1}{2}$.
The exponents associated with the roughening phase transition are
to first order in $\epsilon'$
\begin{equation}
	\alpha =  \frac{\sigma}{2(1-2\sigma)} \epsilon',~~~~ z = 2.
\end{equation}
For $\sigma=0$ the exponents reduce to the well-known results
$\alpha=0$ and $z=2$ for the KPZ equation
(see, e.g., \cite{frey-tauber:1994}).

\section{Conclusions}

\begin{figure}[htb]
\centerline{
	\epsfxsize=7.0cm
	\epsfbox{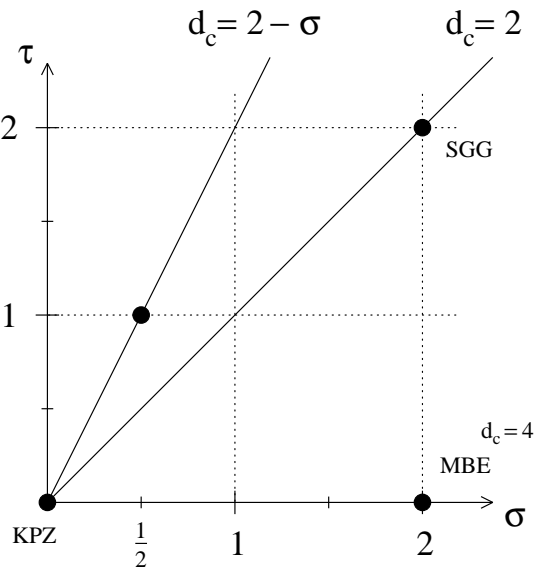}
	\vspace*{0.2cm}
	}
\caption{Universality classes and critical dimensions for the GKE equation.
	Below the line $d_c=2-\sigma$, i.e., for $\tau<2\sigma$, every point
	represents a distinct universality class. Above the line
	$d_c=2-\sigma$, i.e., for $\tau \ge 2\sigma$, every vertical line
	represents a different universality class.
	The KPZ, SGG and MBE models are shown with solid circles.
	The circle at $\sigma=\frac{1}{2}, \tau=1$ divides the
	$\tau=2\sigma$ line into two parts.
	The part with $\sigma \le \frac{1}{2}$
	where the $\epsilon$ expansion does not exist,
	and the part $\sigma>\frac{1}{2}$ where the $\epsilon$
	expansion does exist.
	}
\label{fig:tau-sigma}
\end{figure}

In this paper we have performed a renormalization group
analysis of the GKE growth equation which is an
equation with a generalized conservation law described by an integral kernel.
The results for the FPs and critical exponents
have shown that the GKE equation encompasses a range
of different universality classes, cf.\ Fig.~\ref{fig:tau-sigma}.

For $\tau<2\sigma$, every point represents a distinct universality class,
with the SGG and MBE models belonging to this case.
For all these universality classes we were able to obtain the
critical exponents exactly, and the values are
given in Eqs.~(\ref{eq:gke-exp-a}) and (\ref{eq:gke-exp-z}).
Furthermore, the exponents fulfill the identity (\ref{eq:exp-id-for-gke}).

For $\tau \ge 2\sigma$, every vertical line represents a
different universality class. The KPZ equation belongs to this case.
Moreover,
we noted the breakdown of the $\epsilon$
expansion for $\sigma<\frac{1}{2}$. As a consequence,
estimates for the critical exponents could only be obtained
by a direct substitution of the $g^*$ FP value into the expressions for
the exponents, resulting in the values in
Eqs.~(\ref{eq:a-direct}) and (\ref{eq:z-direct}).
However, for $\sigma>\frac{1}{2}$ the $\epsilon$
expansion could be used to obtain the exponent values
as given in Eqs.~(\ref{eq:a-eps}) and (\ref{eq:z-eps}).

In the GKE equation we have the two free parameters $\tau$
and $\sigma$. As a result, we can always choose these values
in order to obtain agreement with values for $\alpha$ and $z$
determined in an experiment.
However, unless one can argue that the experiment
contains non-local growth effects described by a kernel this does not give
the true explanation of the experiment.

\section*{Acknowledgements}

I acknowledge discussions with Rodolfo Cuerno,
Lu\'{\i}s Amaral and Stefano Zapperi,
and financial support from the Carlsberg Foundation.
The Center for Polymer Studies is supported by NSF.

%\vspace*{-0.3cm}

%\end{references}

\end{document}